\documentclass[12pt]{article}
\usepackage{amsmath,amssymb,bm,enumitem}
\usepackage{epsfig,color,curvesls}
\usepackage{amsfonts}
\usepackage{booktabs} %Table theme
\usepackage{multirow} % for Table
\def\boxit#1{\vbox{\hrule\hbox{\vrule\kern6pt \vbox{\kern6pt#1\kern5pt}
\kern6pt\vrule}\hrule}}

\newenvironment{proof}{\trivlist\item[\hskip \labelsep{\sc Proof:}]}
 {\unskip\nobreak\ \lower.3ex\hbox{$\Box$}\endtrivlist}
\newtheorem{theorem}{Theorem}[section]

\def\boxit#1{\vbox{\hrule\hbox{\vrule\kern6pt
          \vbox{\kern6pt#1\kern6pt}\kern6pt\vrule}\hrule}}
\begin{document}

\def\spacingset#1{\renewcommand{\baselinestretch}%
{#1}\small\normalsize} \spacingset{1}

\title{Empirical Likelihood Based Bayesian Variable Selection} 

 \author{Yichen Cheng\\
 Institute for Insight, Georgia State University\\ 
and\\
 Yichuan Zhao \\
  Department of Mathematics and Statistics, Georgia State University}

\maketitle

\bigskip
\begin{abstract}

Empirical likelihood is a popular nonparametric statistical tool that does not
require any distributional assumptions. 
In this paper, we explore the possibility of conducting variable selection via Bayesian empirical likelihood. 
We show theoretically that when the prior distribution
satisfies certain mild conditions, the corresponding Bayesian empirical
likelihood estimators are posteriorly consistent and variable selection consistent. 
As special cases, we show the prior of Bayesian empirical likelihood LASSO and SCAD satisfies such conditions 
and thus can identify the non-zero elements of the parameters with probability tending to $1$. 
In addition, it is easy to verify that those conditions are met for other widely used priors such as ridge, elastic net and adaptive LASSO.
Empirical likelihood depends on a parameter that needs to be obtained by numerically solving a non-linear equation. 
Thus, there exists no conjugate prior for the posterior distribution, 
which causes the slow convergence of the MCMC sampling algorithm in some cases. 
To solve this problem, we propose a novel approach, which uses an approximation distribution as the proposal. 
The computational results demonstrate quick convergence for the examples used in the paper. 
We use both simulation and real data analyses to illustrate the advantages of the proposed methods.
\end{abstract}

\noindent%
\par
\vfill

\newpage
\spacingset{1.45} % DON'T change the spacing!

\section{Introduction}

Since the introduction of empirical likelihood (EL) (Owen, 1988; 1990), it has been applied to many estimation problems, including the estimation of smooth functions of means (Diciccio et al., 1991),
density (Hall and Owen, 1993; Chen, 1996), copulas (Chen et al., 2009), area under the receiver
operating characteristic (ROC) curve (Qin and Zhou 2006; Yang and Zhao, 2013, 2015; Chrzanowski, 2014), etc. 
Owen (1988) showed that Wilks' theorem holds for the empirical likelihood ratio. 
That is, if we define the empirical likelihood ratio about a $p$-dimensional parameter $\bm{\beta}$ as $R(\bm{\beta})$,
then $R(\bm{\beta}) \rightarrow \chi^2_p$ as $n\rightarrow \infty$ for a fixed $p$.
Thus, the empirical likelihood offers a nice tool for constructing confidence intervals in the nonparametric setting. 
Furthermore, as shown by Diciccio et al. (1991), the empirical likelihood is Bartlett-correctable. 
Thus, the error in the coverage rate for EL based confidence intervels is smaller than that of other nonparametric methods such as the bootstrap.
The advantages of EL include the nonparametric nature, 
the seamless incorporation of prior information and the desirable shape of the confidence region. 
For a more detailed overview of recent development on EL, please refer to Chen and Keilegom (2009).

With the fast developments of data collection techniques, high-dimensional data has become ubiquitous. 
Recent developments of EL also include extending EL to the high-dimensional settings. 
Hjort et al. (2008) showed that when $p=o(n^{1/3})$, $(2p)^{-1/2}\{R(\bm{\beta})-p\}\overset{d}{\rightarrow} \mathcal{N}(0,1)$. 
Chen et al. (2009) further extended the conclusion to $p=o(n^{1/2})$. 
Tang and Leng (2010) proposed a penalized EL for variable selection for linear regression models. 
The proposed method can achieve variable selection, parameter estimation and inference simultaneously. 
Leng and Tang (2012) extended the penalized EL to the general estimating equations setting. 
They showed that the obtained estimator is efficient and has the oracle property 
(Fan and Li, 2011; Fan and Peng, 2004).
Lahiri and Mukhopadhyay (2012) proposed a penalized method for mean estimation based on the adjusted EL. 
Their method is a  modified version of the penalized EL of Bartolucci (2007). 
They showed the resulting method works for $p>n$ cases. 
Chang et al. (2015) developed a generalized EL under moment restrictions for dependent data. 
Recently, Chang et al. (2018) proposed a new scope for penalized EL of general estimating equations. 
They proposed a new penalized EL by applying two penalty functions, which regularize the model parameters and the
associated Lagrange multiplier in the optimizations of EL, respectively. 
They showed that under their setting, both the number of parameters and the number of estimating equations are allowed to grow exponentially with the sample size. 

Accompanying EL and its recent developments (e.g., the jackknife
empirical likelihood by  Jing et al., 2009), EL based Bayesian methods are also proposed.
Lazar (2003) examined the Bayesian counterpart of the empirical likelihood by using
the empirical likelihood in place of a true likelihood to obtain the posterior. 
Results show that, under Bayesian framework, shorter posterior intervals are
achieved with comparable coverage rate. Lazar (2003) also used empirical
evidence to show that the resulting posterior is a valid likelihood
in the sense of Monahan and Boos (1992). She further showed that the
posterior follows asymptotically a normal distribution. 
As another justification of the Bayesian empirical likelihood (BEL),
Schennach (2005) explains the close connection between BEL and a nonparametric Bayesian procedure, thus shows the validity
of an empirical-likelihood-type likelihood. 
Since then, BEL has been successfully applied to problems such as quantile regression
Yang and He (2012) and Bayes model selection (Chib et al. 2018).
As another example, Lancaster and Jun (2010) proposed the Bayesian quantile regression based on Schennach's work.
Very recently, Bedoui and Lazar (2020) and Moon and Bedoui (2020) proposed Bayesian empirical likelihoodhood based varaible selection under the ridge, LASSO and elastic net prior. However, their work is more focused on the computational aspect.

Consider the linear regression setting. Let ${\bf y}$ be the outcome, which
is a vector of length $n$. Let ${\bf X}$ be the covariate matrix, which
is a $n\times p$ matrix with each column representing a covariate.
We assume that ${\bf X}$ and ${\bf y}$ follow a linear regression
model with the distribution of the error term ${\bm \epsilon}$ unspecified:
$
{\bf y}={\bf X}{\bm \beta}+{\bm \epsilon}, E(\epsilon_i ) = 0,
$
where $\epsilon_i$ is the $i$th element of $\bm \epsilon$.

An empirical likelihood method works by finding the ${\bm \beta}$ that maximizes the empirical likelihood
function defined below:
$
L({\bm \beta})=\sup\{\prod_{i=1}^{n}w_{i}:w_{i}\geq0,\sum_{i=1}^{n}w_{i}=1,\sum_{i=1}^{n}w_{i}{\bf U}_{i}({\bm \beta})=0\},
$
where ${\bf U}_{i}({\bm \beta})={\bf x}_{i}(y_{i}-{\bf x}_{i}^{T}{\bm \beta})$, $y_{i}$ is the
$i$th element of ${\bf y}$ and ${\bf x_{i}}$ is the transpose
of the $i$th row of ${\bf X}$. By the Lagrange multiplier method, 
$
w_{i}=n^{-1}\{1+{\bm \lambda}_{\bm \beta}^{T}{\bf U}_{i}({\bm \beta})\}^{-1},
$
where ${\bm \lambda}$ solves the equation 
$
\sum_{i=1}^{n}  {\bf U}_{i}({\bm \beta})/\{1+{\bm \lambda}_{\bm \beta}^{T}{\bf U}_{i}({\bm \beta})\}=0.
$

Under the situation where the number of covariates is large, it may
be reasonable to assume only a subset of the covariates play a role
in the model. Under such an assumption, similar to the penalized method
proposed for the regular linear regression model, Tang and Leng (2010)
proposed a penalized empirical likelihood to enforce the sparsity
for ${\bm \beta}$. In summary, the penalized empirical likelihood estimator
$\hat{\bm \beta}$ is defined as the minimizer of 
$
l_{P}(\bm \beta)=\sum_{i=1}^{n}\log\{1+{\bm \lambda}_{\bm \beta}^{T}{\bf U}_{i}(\bm \beta)\}+n\sum_{j=1}^{p}P(|\beta_{j}|),
$
where the function $P$ is some penalty function. 
To show the oracle property of the proposed method, they used the smoothly clipped absolute deviation penalty (SCAD). 
They showed the oracle property holds for both the mean vector estimation and the linear regression models. 

In this paper, we propose a class of Bayesian empirical likelihood (BEL) based variable selection methods, including BEL LASSO and BEL SCAD as special cases.
We discuss the posterior consistency 
and variable selection consistency 
for the mean
vector estimation and the linear regression problem.
We show a group of prior distribution will lead to consistent posterior distributions. 
That is, the posterior distribution of each parameter will go to the true value with probability tending to $1$ as the sample size increases. 
As a result, the correct variables will be identified with probability close to $1$.

The rest of the paper is organized as follows. 
Section 2 introduces BEL based variable selection. 
The posterior consistency and variable selection consistency results are presented in Section 3. 
In Section 4, we address several computational issues.
In Section 5, we carry out extensive simulation studies. 
In Section 6, we use blue real data examples to illustrate the proposed BEL approach.
Section 7 concludes this paper with a discussion about the connection between
our method and other existing approaches.

\section{BEL for variable selection}

\subsection{BEL LASSO for the mean vector ${\mu}$}

Assume we have a collection of $n$ independent multivariate random vectors
$\{{\bf x}_{i},i=1,\ldots,n,\,{\bf x}_{i}\in\mathcal{R}^{p}\}$ with $E({\bf x}_{i})={\bm \mu}_{0}$ and $var({\bf x}_{i})={\bm \Sigma}$.
The dimension $p$ is allowed to grow with the sample size, $n$.
 The empirical likelihood for ${\bm \mu}$ is:
$
L({\bm \mu})=\sup\{\prod_{i=1}^{n}w_{i}:w_{i}\geq0,\sum_{i=1}^{n}w_{i}=1,\sum_{i=1}^{n}w_{i}({\bf x}_{i}-{\bm \mu})=0\}.
$
By the Lagrange multiplier method, 
$L({\bm \mu})=\{\prod_{i=1}^{n} w_{i}:w_{i}=n^{-1}\{1+{\bm \lambda}_{\mu}^{T}({\bf x}_{i}-{\bm \mu})\}^{-1}\}$,
where ${\bm \lambda}_{\mu}$ solves the equation 
$
\sum_{i=1}^{n} ({\bf x}_{i}-{\bm \mu})/\{1+{\bm \lambda}_{\mu}^{T}({\bf x}_{i}-{\bm \mu})\}=0.
$

To enforce the sparsity for $\bm \mu$,
we use the same penalty term as used in Bayesian LASSO (Tibshirani, 1996). 
Then, the penalized empirical likelihood can be written as:
$
l_{p}(\bm \mu)=\sum_{i=1}^{n}\log\{1+{\bm \lambda}_{\bm \mu}^{T}({\bf x}_{i}-{\bm \mu})\}+\gamma\sum_{j=1}^{p}|\mu_{j}|,
$

As noted in Tibshirani (1996) and Park and Casella (2008), the LASSO
estimates can be interpreted as the posterior mode estimates when
the parameters follow independent and identical Laplace priors. 
Further, the Laplace prior can be viewed as a mixture of normal distributions. 
Motivated by this observation, we propose a Bayesian framework based on the penalized empirical likelihood. 
The priors are specified as follows:
$ {\bm \mu} \sim \mathcal{N}({\bf 0},{\bf D}_{\tau})$, and 
$\pi(\tau_1^2,\ldots,\tau_p^2)  = \prod_{j=1}^p \frac{\gamma^2}{2} \exp(-\gamma^2 \tau_j^2/2)$,
where ${\bf D}_{\tau} = diag(\tau_1^2,\ldots,\tau_p^2)$. 

Therefore, the posterior is
\begin{equation*}
p({\bm \mu}, {\bf D}_{\tau}|X,{\bf y}) \propto \prod_{i=1}^{n}\frac{1}{\{1+{\bm \lambda}_{\mu}^{T}({\bf x}_i - {\bm \mu})\}}\prod_{j=1}^p (2\pi\tau_j^2)^{-1/2}  \exp\left( -\frac{\mu_j^2}{2\tau_j^2} \right) \prod_{j=1}^p \frac{\gamma^2}{2} \exp(-\gamma^2 \tau_j^2/2),
\end{equation*}
where ${\bm \lambda}_{\mu}$ solves the equation 
$
\sum_{i=1}^{n} ({\bf x}_{i}-{\bm \mu})/\{1+{\bm \lambda}_{\mu}^{T}({\bf x}_{i}-{\bm \mu})\}=0.
$

\subsection{BEL LASSO for linear models}

Similar to the mean vector situation, we can use an $L_{1}$ penalty
term to enforce the sparsity of ${\bm \beta}$ in the linear regression
model. 
We assume that ${\bf X}$ and ${\bf y}$ follow a linear regression model with the distribution of the error term ${\bm \epsilon}$ unspecified:
$
{\bf y}={\bf X}{\bm \beta}+{\bm \epsilon}, E({\bm \epsilon})={\bf 0}.
$
Without loss of generality, we assume that both ${\bf y}$ and ${\bf X}$ are
centered to have mean zero such that $E({\bf y})={\bf 0}$, $E({\bm x}_{i})=0$ and
$var({\bf x}_{i})={\bm \Sigma}$. We also assume that the error terms $\epsilon_{i}$'s
are $i.i.d.$ with mean zero. 
The penalized empirical likelihood can be written as
$
l_{p}({\bm \beta})=\sum_{i=1}^{n}\log\{1+{\bm \lambda}_{\bm \beta}^{T}{\bf U}_{i}({\bm \beta})\}+\gamma\sum_{j=1}^{p}|\beta_{j}|,
$
where ${\bf U}_{i}({\bm \beta}) = {\bf x}_i(y_i - {\bf x}_i^T {\bm \beta})$.

Similar to the mean vector case,  the priors are set as follows:
$ {\bm \beta} \sim \mathcal{N}({\bf 0},{\bf D}_{\tau})$, and
$\pi(\tau_1^2,\ldots,\tau_p^2)  \sim \prod_{j=1}^p \frac{\gamma^2}{2} \exp(-\gamma^2 \tau_j^2/2)$,
where ${\bf D}_{\tau} = diag(\tau_1^2,\ldots,\tau_p^2)$.

Then, the posterior likelihood is
\begin{equation*}
p({\bm \beta}, {\bf D}_{\tau}|X, {\bf y}) \propto \prod_{i=1}^{n}\frac{1}{1+{\bm \lambda}_{\bm \beta}^{T}{\bf U}_{i}({\bm \beta})}\prod_{j=1}^p (2\pi\tau_j^2)^{-1/2}  \exp\left( \frac{\beta_j^2}{2\tau_j^2} \right) \prod_{j=1}^p \frac{\gamma^2}{2} \exp(-\gamma^2 \tau_j^2/2),
\end{equation*}
where ${\bm \lambda}_{\bm \beta}$ solves the following equations 
$
\sum_{i=1}^{n} {\bf U}_{i}({\bm \beta})/\{1+{\bm \lambda}_{\bm \beta}^{T}{\bf U}_{i}({\bm \beta})\}=0.
$

\subsection{From LASSO to SCAD priors}

The LASSO prior is easy to implement and the posterior consistency and variable selection consistency  can be established. 
Fan and Li (2011) argue that compared to the LASSO penalty, SCAD enjoyes more desired properties. 
Thus, we consider the SCAD prior as an alternative to the LASSO prior, which is described below. 

The SCAD penality (Fan and Li, 2011) $P_{\gamma}$ is defined as $P_{\gamma}(0) = 0$ and 
$P'_{\gamma}(\theta) = \gamma \{ I(\theta<\gamma)+\frac{(a\gamma - \theta)_+}{(a-1)\gamma} I(\theta > \gamma) \}$, 
with $a>2$ being some fixed number and $\theta>0$.
In Fan and Li (2001), they recoomend to set $a=3.7$. We fix $a$ to be $3.7$, throughout the paper.
Thus, a SCAD prior can be defined as $\pi(\bm \beta) \propto \exp\{-\sum_j P_{\gamma}(|\beta_j|)\}$, 
which is not in the form of a known distribution. 
Motivated by Li (2011), we approximate the penalty function by a linear Taylor expansion such that:
$P_{\gamma}(\theta) = P_{\gamma}(\theta^{0}) +  P'_{\gamma}(\theta^{0}) (\theta - \theta^{0}) $.
Then, at the $t$th iteration of the MCMC process,  
the prior distribution can be approximated by $\pi(\bm \beta) \propto \prod \exp\{- P'_{\gamma}(\beta_j^{(t-1)}) |\beta_j|\}$,
in a way similar to what is done with the LASSO prior.

%%%%%%%%%%%%%%%%%%%%%%%%%%%%%%%%%%%%%%%%%%%%%%%%%%%%%%%%%%%%%%%%%%%%%%%%%%%%%%%%%%%%%%%%%%%%%%%%%%%%%%%%%%%%%%%%%%%%%%%%%%%%

%\lhead[\footnotesize\thepage\fancyplain{}\leftmark]{}\rhead[]{\fancyplain{}\rightmark\footnotesize\thepage}%Put this line in Page 2

{ \section{Posterior consistency and variable selection consistency}}

In this section, we show the posterior consistency and the variable selection consistency. 
The results for the mean vector ${\bm \mu}$ are given in Section 3.1 and those for the ${\bm \beta}$ in the linear regression model
in Section 3.2. 
We first discuss the sufficient conditions for the consistencies. 
We then show the proposed BEL LASSO satisfies those conditions, and thus provides consistent results. 

From Figure 1 in Fan and Li (2011), it is easy to observe that, compared with the LASSO penalty, SCAD penalty gives the same level of penalty when $|\theta|\leq \gamma$, 
while puts less penalty when $|\theta|>\gamma$, where $\gamma$ is the tuning parameter. 
If the regularity conditions on the prior are satisfied for the LASSO prior, it is straightforward to show that those are also satisfied for the SCAD prior.  
Thus, in this section, we only show the regularity conditions are satisfied for the LASSO prior. 
The proof for SCAD prior is very similar, and thus omitted. The proofs for the theorems presented in this Section can be found in the Online Supplementary Material.

\subsection{Posterior consistency and variable selection consistency for the mean vector}

In this subsection, we present the consistency results for of the mean vector estimator. 
The main results are presented in Theorems \ref{thm:mean} and \ref{thm:mean_selection}. 
In order to show the posterior consistency of the mean estimators, some regularity conditions are required. 
\begin{description}
\item [Condition 1] Let $\sigma_{1}$ ($\sigma_{p}$) be the smallest
(largest) singular values of ${\bm \Sigma}\triangleq cov({\bf x})$, respectively. 
Then there exist constants $L$ and $U$, such that
 $0<L<\liminf\sigma_{1}<\limsup\sigma_{p}<U<\infty$. 
\item [Condition 2] $E_{{\bm \mu}}|X_{ij}|^{k}<\infty$ for some $k\geq3$. 
\end{description}
Under Conditions 1 and 2, the posterior consistency of ${\bm \mu}$
holds if the prior distribution assigns a non-trivial probability to the neighborhood of the true value ${\bm \mu_0}$. 
The results are formally presented in Theorem \ref{thm:mean}.

\begin{theorem} \label{thm:mean}
(Posterior consistency) Define $e_n = (p/n)^{1/2-\delta}$, 
$\mathcal{D}_{{\bm \mu}} = \{{\bm \mu}:\| {\bm \mu} - {\bm \mu}_0\| \leq c_1 e_n \}$ with $\delta, c_1 >0$ 
and $\delta$ satisfies $p^{1-\delta-1/4k}/n^{1/2-\delta-1/4k} \rightarrow 0$ as $n\rightarrow 0$. 
Define $\mathcal{B}_{{\bm \mu}} = \{{\bm \mu}:\| {\bm \mu} - {\bm \mu}_0\| > c_1 e_n \}$.
Then under Conditions 1 and 2, the posterior of ${\bm \mu}$ under prior $\Pi_{n}({\bm \mu})$ is strongly consistent. 
That is, $\Pi(\mathcal{B}_{{\bm \mu}}|{\bf X})=\Pi({\bm \mu} : \|{\bm \mu}-{\bm \mu}_{0}\|>c_1 e_{n}|{\bf X})\rightarrow0$
$Pr_{{\bm \mu}_{0}}$ almost surely if the prior satisfies
$
\Pi \left({\bm \mu}:\|{\bm \mu} - {\bm \mu}_{0}\|^2 < a_{n} U \right)>\exp(-d_{n}n),
$
for $a_{n}+d_{n}<b_{n}$ and $b_{n} = c_1^2 e_{n}^{2}/(4U)$. 
\end{theorem}

The above theorem shows that the posterior samples obtained from the proposed BEL methods under a group of priors will be close enough to the truth. 
However, as a general Bayesian method with shrinkage prior, the BEL based methods do not force any of the estimate to be exactly zero. 
Motivated by the results obtained in Theorem \ref{thm:mean}, we can use a hard-thresholded version of the Bayesian samples to achieve the sparsity: 
$\tilde{\mu}_j \triangleq \mu_jI(|\mu_j|>c_1e_n)$. 
The following theorem shows that under a mild condition on the magnitude of the true non-zero means, the variable selection consistency can be achieved.

\begin{theorem} \label{thm:mean_selection}
(Variable selection consistency) Under the conditions of Theorem \ref{thm:mean}, and further assume that $min_{j\in \mathcal{A}_\mu}|\mu_j|>2c_1e_n$.
We have $\Pi\{I(\tilde{\mu}_j=0) = I(\mu_{0,j}=0) \mbox{\ for\ all\ } j|{\bf X}\} \rightarrow 1$ $Pr_{\bm \mu_0}$ almost surely,
where $\bm \mu_0$ is the true value of the mean vector. 
That is, we can consistently select the non-zero elements.
\end{theorem}

\iffalse
\begin{proof}
From Theorem  \ref{thm:mean}, we have $\Pi({\bm \mu} : \|{\bm \mu}-{\bm \mu}_{0}\|>c_1 e_{n}|{\bf X})\rightarrow0$. 
Thus, it is straightforward to have: $\Pi( |\tilde{\mu}_j-\mu_{j0}|>c_1 e_{n} \mbox{\ for\ at\ least\ one\ }  j|{\bf X})\rightarrow0$.
Since $min_{j\in \mathcal{A}_\mu}|\mu_j|>2c_1e_n$, we have $\Pi\{I(\tilde{\mu}_j==0) \neq I(\tilde{\mu}_j) \mbox{\ for\ at\ least \  one } j|{\bf X}\} \rightarrow 0$.
\end{proof}
\fi

Note that $e_n = o(1)$. The condition on the magnitude of the true non-zero means can be easily satisfied in practice. 
The previous two theorems show the validity of a group of priors under the BEL setting. 
The following theorem shows, as a special case, the Laplace prior satisfies the conditions specified in Theorem \ref{thm:mean}. 
As a result, both the posterior consistency and the variable selection consistency of BEL LASSO are established. 

\begin{theorem} \label{thm:3.2}
Let $\mathcal{A}_{\mu} = \{j:\mu_{0,j} \neq 0\}$ and $s_1=|\mathcal{A}_\mu|$, which is the number of non-zero element of ${\bm \mu}_0$ and assumed to be a fixed finite number. 
Then, under Conditions 1-2, and the conditions specified in Theorem \ref{thm:mean} about $e_n$,
the Laplace prior $f(\mu_{j}|\gamma)=(\gamma/2)\exp(-\gamma|\mu_{j}|)$
with scale parameters $\gamma$ yields a strongly consistent posterior
if $\gamma=O\{ (p/n)^{1-h} n \}$ and $0<h<2\delta$.
Furthermore, if $min_{j\in \mathcal{A}_\mu}|\mu_j|>2c_1e_n$,
we have $\Pi\{I(\tilde{\mu}_j=0) = I(\mu_{0,j}=0)  \mbox{\ for\ all\ } j|{\bf X}\} \rightarrow 1$ $Pr_{\bm \mu_0}$ almost surely. 
\end{theorem}

Note, the condition that $s_1$ is fixed is only needed for Theorem \ref{thm:3.2}, 
not for Theorems \ref{thm:mean} and \ref{thm:mean_selection}. 
Under the condition that the number of non-zero element is fixed and finite, we can make use of the extra information and use a Laplace prior that put the prior mass for a large proportion
of the elements to be close to zero. 
Thus, it ensures the consistency in the special case of sparsity. Further, note that $\gamma/n$ in our setting is working in a similar way as the penalty term in the frequentist setting. $\gamma/n$ approaches zero as the sample size increases, but at a slower rate than the estimates for the zero elements, thus, working as a threshold to differentiate the non-zero elements from the zeros.

\subsection{Posterior consistency and variable selection consistency  for ${\beta}$ in the linear regression model}

In order to show the posterior consistency of ${\bm \beta}$, we need some additional conditions on the error terms. 
\begin{description}
\item[Condition 3]: The random errors $\{\epsilon_{i}\}_{i=1}^{n}$ are
independently and identically distributed with $E(\epsilon_{i})=0$
and $E(|\epsilon_{i}|^{3})<\infty$.
\end{description}

\begin{theorem} \label{thm:3}
(Posterior consistency)
Define $e_n = (p/n)^{1/2-\delta}$ with $\delta>0$ and $\delta$ satisfies that $p^{1-\delta-1/4k}/n^{1/2-\delta-1/4k} \rightarrow 0$ as $n\rightarrow 0$. 
Under Conditions 1-3, the posterior of ${\bm \beta}$ under prior $\Pi({\bm \beta})$
is strongly consistent, that is, 
$\Pi(\mathcal{B}_{{\bm \beta}}|{\bf X})=\Pi_{n}({\bm \beta}:\|{\bm \beta} - {\bm \beta}_{0}\|>c_4 e_{n}|X)\rightarrow0$,
$Pr_{{\bm \beta}_{0}}$ almost surely if the prior satisfies
$
\Pi\left({\bm \beta}:\|{\bm \beta} - {\bm \beta}_{0}\|^2<a_{n}U\right)>\exp(-d_{n}n),
$
for $a_{n} + d_{n}<b_{n}$ and $b_{n}=c_5^2e_{n}^{2}/(4U)$, where ${\bm \beta}_{0}$ denote the truth. 
\end{theorem}

\begin{theorem} \label{thm:beta_selection}
(Variable selection consistency) Under the conditions of Theorem \ref{thm:3}, and further assume that $min_{j\in \mathcal{A}_\beta}|\beta_j|>2c_4e_n$, 
where $\mathcal{A}_\beta = \{j:\beta_{j0}\neq 0\}$.
We have $\Pi\{I(\tilde{\beta}_j=0) = I({\beta}_{0,j}=0) \mbox{\ for\ all\ } j|{\bf X},{\bf y}\} \rightarrow 1$ $Pr_{\bm \beta_0}$ almost surely. That is, we can consistently select the non-zero variables.
\end{theorem}

The proof of Theorem \ref{thm:beta_selection} is similar to that of Theorem \ref{thm:mean_selection} and is thus omitted.

\begin{theorem} \label{thm:4}
Let $\mathcal{A}_{\beta} = \{j:\beta_{0,j} \neq 0\}$ and $s_2=|\mathcal{A}_\beta|$, which is the number of non-zero element of ${\bm \beta}_0$ and assumed to be a fixed finite number. 
Define $e_n = (p/n)^{1/2-\delta}$ with $\delta>0$ and $\delta$ satisfies $p^{1-\delta-1/4k}/n^{1/2-\delta-1/4k} \rightarrow 0$ as $n\rightarrow 0$. 
Then, under Conditions 1-3, the Laplace prior $f(\beta_{j}|\gamma)=(\gamma/2)\exp(-\gamma|\beta_{j}|)$ with scale parameters $\gamma$ yields a strongly consistent posterior
if $\gamma=O\{ (p/n)^{1-h} n \}$ and $0<h<2\delta$.
Furthermore, if $min_{j\in \mathcal{A}_\beta}|\beta_j|>2c_4e_n$,
we have $\Pi\{I(\tilde{\beta}_j=0) = I({\beta}_{0,j}=0)  \mbox{\ for\ all\ } j|{\bf X},{\bf y}\} \rightarrow 1$ $Pr_{\bm \mu_0}$ almost surely.
\end{theorem}

The proof of Theorem \ref{thm:4} follows a similar argument as that of Theorem \ref{thm:3.2}, and thus is omitted.

{ \section{Computational issues}}
In this section, we discuss the selection of the hyperparameter and the computational issue related with sampling from the posterior distribution.

\subsection{Selection of hyperparameter $\gamma$}
The proposed BEL LASSO has a hyperparameter that needs to be specified. To select the value of the hyperparameter $\gamma$, we have two different approaches:
1) Using empirical Bayes approach by treating $\gamma$ as missing values and estimate $\gamma$ using EM algorithm.
2) Specify a prior on $\gamma$.
Next, we discuss in detail about the implementation of the two options.

\subsubsection{Empirical Bayes and EM algorithm}

For illustration purpose, we use the mean vector as an example, the procedure for the linear regression model follows similarly. 
As a reminder, the full likelihood $p({\bf X},{\bm \mu},{\bf D}_{\tau},\gamma)$ is proportional to:
\begin{equation*}
\prod_{i=1}^n \frac{1}{\{1+{\bm \lambda}_{\mu}^T({\bm X_i} - {\bm \mu})\}} \prod_{j=1}^p (2\pi\tau_j^2)^{-1/2}  \exp\left( \frac{\mu_j^2}{2\tau_j^2} \right) \prod_{j=1}^p \frac{\gamma^2}{2} \exp(-\gamma^2 \tau_j^2/2).
\end{equation*}

Applying the EM algorithm, we get at the $k$th iteration:
\begin{equation} \label{eq:EM}
Q(\gamma|\gamma^{(k-1)}) = p \ln(\gamma^2) - \frac{\gamma^2}{2} \sum_{j=1}^p E_{\lambda^{(k-1)}}(\tau_j^2|\bf{X}).
\end{equation}

By maximizing $\gamma$ for Equation (\ref{eq:EM}), one has
$
\gamma^{(k)} = \sqrt{ 2p/\{\sum_{j=1}^p E_{\gamma^{(k-1)}}(\tau_j^2|{\bf X})\}},
$
where $E_{\lambda^{(k-1)}}(\tau_j^2|{\bf X})$ can be estimated using the samples obtained up to iteration $k-1$. 

Such a Monte Carlo based EM algorithm was first introduced by Casella (2001). 
It should be noted that such an algorithm almost adds no computational burden to the original sampling methods, since the expectation can be estimated using existing Monte Carlo samples at each iteration.

\subsubsection{Priors for hyperparameter $\gamma$}

Instead of treating $\gamma$ as a missing value and estimating it using an EM algorithm, 
an alternative is to specify a prior for $\gamma$ and let the data decide the distribution of $\gamma$. 
We follow Park and Casella (2008) and use the class of Gamma prior on $\gamma^2$ of the form:
\[
\pi(\gamma^2) = \frac{\delta^r}{\Gamma(r)}(\gamma^2)^{r-1} \exp(-\delta' \gamma^2),
\]
where $r>0$ and $\delta'>0$.

The Gamma prior on $\gamma^2$ allows for the conjugacy on $\gamma^2$, and thus has a better sampling mixing. 
Under such a prior specification, the posterior of $\gamma^2$ follows a Gamma distribution with shape parameter $p+r$
and rate parameter $\sum \tau_j^2/2+\delta'$. 
The posteriors for other parameters keep unchanged.

\subsection{Sampling from the posterior distribution}
Under both the settings described in the previous subsection, it is easy to verify that the posterior of $1/\tau_j^2$ follows
an inverse-Gaussian distribution, with $\lambda' = \gamma^2$ and $\mu' = \sqrt{\gamma^2/\mu_j^2}$. 
However, the posteriors of ${\bm \mu}$ and ${\bm \beta}$ do not follow any known distribution. 
We discuss two promising approaches in this subsection. 
For simplicity, we describe the procedure for sampling from the posterior distribution of ${\bm \mu}$. 
The procedure to sample from the posterior distribution of ${\bm \beta}$ is similar.

\subsubsection{Gibbs sampling with embedded Metropolis-Hasting algorithm}

\noindent {\bf Algorithm 1:}

Initialization: Set ${\bm \mu}^{(1)} = \bar{\bf X}$, 
$(\tau_j^{(1)})^2 = n/\sum_i (X_{ij} - \mu^{(1)}_j)^2$, 
$\gamma^{(1)}=\sqrt{2 \sum_{ij} (X_{ij} - \mu_j^{(1)})^2/n}$. 

At iteration $k$:
\begin{itemize}
\item Update $\gamma$: 
\[
\gamma^{(k)} = \sqrt{\frac{2p}{\sum_{j=1}^p \hat{E}_{\lambda^{(k-1)}}(\tau_j^2|{\bf X})}},
\]
where $\hat{E}_{\lambda^{(k-1)}}(\tau_j^2|{\bf X}) = \sum_{t=1}^{k-1}\left(\tau_j^{(t)}\right)^2/(k-1)$.
\item Update $\tau_j^2$ for $j =1, \ldots, p$: sample $1/\tau_j^2$ from an inverse Gamma distribution with $\mu' = \gamma^{(k)}/|\mu_j^{(k-1)}|$ and $\lambda' = \left(\gamma^{(k)}\right)^2$.
\item Update ${\bm \mu}$ using Metropolis-Hasting algorithm:
\begin{itemize}
\item[-] Generate $un \sim Unif(0,1)$, $l_1 \sim sample(1:p)$ and $l_2 \sim sample(1:p)$, $l_1 \neq l_2$.
\item[-] Proposal:
\begin{itemize}
\item[1.] If $un \in (0,0.4)$, set ${\bm \mu}^{(k-1/2)} = {\bm \mu}^{(k-1)} + {\bf e}\times rnorm(0,1) \times s$, 
where ${\bf e}$ has norm one and generated uniformly from a $p$-dimensional unit circle, 
$s$ is a user defined parameter that controls the step size and $rnorm(0,1)$ is a random sample from a normal distribution with mean $0$ and variance $1$. 
\item[2.] If $un \in (0.4,0.7)$, set $\mu^{(k-1/2)}_j = \mu^{(k-1)}_j$ for $j \neq l_1$ and $ \mu^{(k-1/2)}_{l_1} = \mu^{(k-1)}_{l_1} + rnorm(0,1) \times s$.
\item[3.] If $un \in (0.7,1)$, set $\mu^{(k-1/2)}_j = \mu^{(k-1)}_j$ for $j \neq l_1$ or $l_2$, $ \mu^{(k-1/2)}_{l_1} = \mu^{(k-1)}_{l_1} + rnorm(0,1) \times s$ and $\mu^{(k-1/2)}_{l_2} =  \mu^{(k-1)}_{l_2} + rnorm(0,1) \times s$.
\end{itemize}
\item[-] The acceptance rate is calculated as:
\[
r = min\left( 1, \frac{p({\bm \mu}^{(k-1/2)}|{\bf X},{\bf D}_{\tau}^{(k)},\gamma^{(k)})}{p({\bm \mu}^{(k-1)}|{\bf X},{\bf D}_{\tau}^{(k)},\gamma^{(k)})} \right),
\]
\item[-] Generate $uni \sim Unif(0,1)$, set ${\bm \mu}^{(k)} = {\bm \mu}^{(k-1/2)}$ if $uni<r$ and ${\bm \mu}^{(k)} = {\bm \mu}^{(k-1)}$ if $uni>r$.
\end{itemize}
\end{itemize}

When the dimension $p$ is large, sampling from the posterior using the algorithm described above can be challenging. When we set $s$ to be large, the acceptance rate can be low and the Monte Carlo samples can get trapped in a local maximum of the posterior. When we set the $s$ to be small, the acceptance rate will increase. 
However, the small movement from one iteration to the next might lead to strong auto-correlation between samples. Also, small $s$ might cause the algorithm to fail to explore the whole posterior sample space. 

\subsubsection{Sampling based on a multivariate normal approximation}
To overcome the difficulty mentioned in the previous subsection, 
we propose a method similar to the idea of Laplace approximation. 
The key idea is to note (by Lemma 1.2 in the Supplemenatry Material) that the empirical likelihood can be approximated by a multivariate normal distribution. 
If we are willing to assume the approximation is accurate, then the posterior distribution of ${\bm \mu}$ given everything else can be approximated by another multivariate normal distribution. Thus, a full Gibbs sampler can be used for sampling both ${\bm \mu}$ and ${\bf D}_{\tau}$. However, as shown in Lemma 1.2, the approximation is only derived for a neighborhood of ${\bm \mu}_0$. 
Thus, the approximation may be inaccurate for ${\bm \mu}$ that is far away from ${\bm \mu}_0$. 
Therefore, we propose to use the approximated posterior as the proposal distribution in the Metropolis-Hasting algorithm. 
By using such an approximation, the mixing of posterior samples will be much better than those obtained from the algorithm introduced in the previous subsection. 
Motivated by these observations, we propose the following approximation-based Gibbs sampler.

\noindent {\bf Algorithm 2:}

Initialization: Set ${\bm \mu}^{(1)} = \bar{\bf X}$, 
$\left(\tau_j^{(1)}\right)^2 = n/\sum_i \left(X_{ij} - \mu^{(1)}_j\right)^2$, 
$\gamma^{(1)}=\sqrt{2 \sum_{ij} \left(X_{ij} - \mu_j^{(1)}\right)^2/n}$.

At iteration $k$:
\begin{itemize}
\item Update $\gamma$: 
\[
\gamma^{(k)} = \sqrt{\frac{2p}{\sum_{j=1}^p \hat{E}_{\lambda^{(k-1)}}(\tau_j^2|{\bf X})}},
\]
where $\hat{E}_{\lambda^{(k-1)}}(\tau_j^2|{\bf X}) = \sum_{t=1}^{k-1}\left(\tau_j^{(t)}\right)^2/(k-1)$.
\item Update $\tau_j^2$ for $j =1, \ldots, p$: sample $1/\tau_j^2$ from an inverse Gamma distribution with $\mu' = \gamma^{(k)}/|\mu_j^{(k-1)}|$ and $\lambda' = \left(\gamma^{(k)}\right)^2$.
\item Update ${\bm \mu}$ by using the Metropolis-Hasting algorithm with a multivariate normal distribution as the proposal distribution:
\begin{itemize}
\item[-] Generate $un \sim Unif(0,1)$, $l_1 \sim sample(1:p)$ and $l_2 \sim sample(1:p)$, $l_1 \neq l_2$.
\item[-] Proposal:
\begin{itemize}
\item[1.] Generate ${\bm \mu}^{(k-1/2)}$ from a multivariate normal distribution with mean $\tilde{{\bm \mu}}$ and covariance matrix $\tilde{\Sigma}$, 
where $\tilde{\bm \Sigma} = (n{\bm \Sigma}_n^{-1}+{\bf D}_{\tau}^{-1})^{-1}$, 
$\tilde{{\bm \mu}} = \bar{X}^T {\bm \Sigma}_n^{-1} \tilde{\bm \Sigma} $, and
${\bm \Sigma}_n = \sum_{i=1}^n ({\bf x}_i - \bar{\bf X})({\bf x}_i - \bar{\bf X})^T/n$.
\item[-] The acceptance rate is calculated as:
\[
r = min\left( 1, \frac{p({\bm \mu}^{(k-1/2)}|{\bf X},{\bf D}_{\tau}^{(k)},\gamma^{(k)})g({\bm \mu}^{(k-1)}|\tilde{{\bm \mu}},\tilde{\bm \Sigma})}{p({\bm \mu}^{(k-1)}|{\bf X},{\bf D}_{\tau}^{(k)},\gamma^{(k)}),g({\bm \mu}^{(k-1/2)}|\tilde{{\bm \mu}}, \tilde{\bm \Sigma})} \right),
\]
\item[-] Generate $uni \sim Unif(0,1)$, set ${\bm \mu}^{(k)} = {\bm \mu}^{(k-1/2)}$ if $uni<r$ and ${\bm \mu}^{(k)} = {\bm \mu}^{(k-1)}$ if $uni>r$.
\end{itemize}
\end{itemize}
\end{itemize}

\begin{figure}[htbp] 
\begin{center}
\begin{tabular}{c}
\epsfig{figure=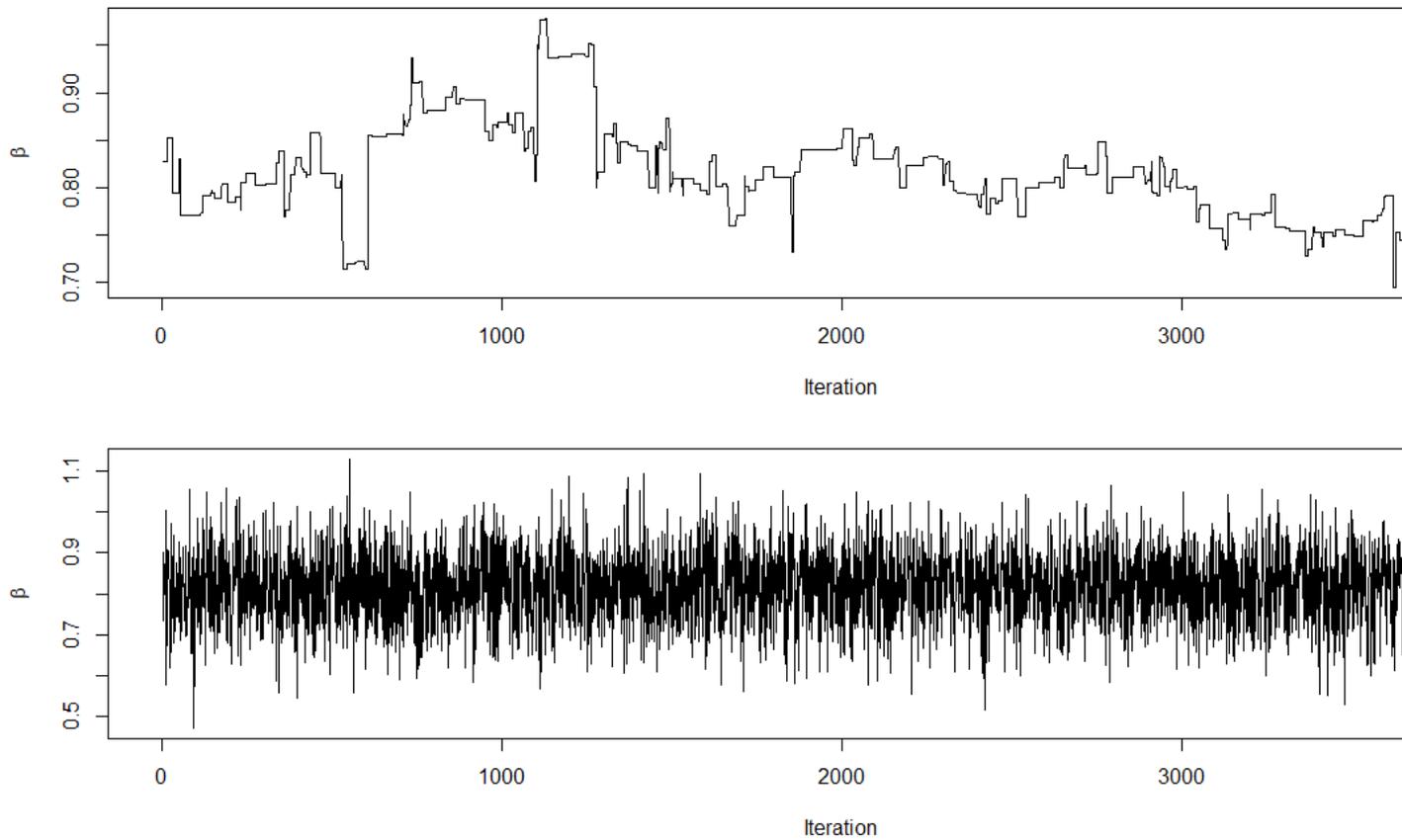,scale = .5} 
\end{tabular}
\caption{The sampling path of $\beta_1$. Top panel: first $5000$ samples obtained using Algorithm 1. Bottom panel: first $5000$ samples obtained using Algorithm 2. }
\label{fig:sample_path}
\end{center}
\end{figure}

To compare the performance of the sampling method proposed in this subsection and the previous subsection, 
we generate a simulated data set with $n=100$ and $p=20$. 
For each element of ${\bf X}$, we generate $X_{ij}$ from the standard normal distribution, $i=1,\ldots,n$, $j=1,\ldots,p$. We let $y = x_1 + 2x_2+3x_3+ 4x_4+5x_5+\epsilon$, with $\epsilon$ following a standard normal distribution. 
We show the sampling path of one parameter in Figure \ref{fig:sample_path}. 
The sampling path of $\beta_1$ using Algorithm 1 is displayed in the top panel, and that of Algorithm 2 is in the bottom panel. As can be seen from the plots, the mixing of Algorithm 2 is much better than that of Algorithm 1.

\section{Simulation study}

\subsection{Simulation for mean vector ${\mu}$}

In this subsection, we compare the performance of the proposed method
with existing approaches for estimating the mean vector ${\bm \mu}$. 
We generate the ${\bf X}$ matrix by assuming ${\bf x}_i={\bm \mu}_{0}+{\bm \Gamma}^{1/2}({\bf z}_i-1)$,
where ${\bm \mu}_{0}=(3,2,1,0.6,0.3,0,0,\ldots,0)$ is a vector of length $p$, ${\bf z}_i$ is a vector of length $p$ with each element 
following a chi-squared distribution with degree of freedom $1$ and
${\bm \Gamma}$ is a matrix with 1 on the diagonal and
$\rho$ everywhere else. 
We consider two situations for $\rho=0.3$ and $\rho=0.7$. 
We consider different combinations of $n$ and $p$, with $n$ being $100$, $200$ or $500$ and $p$ being $10$ or $20$.

\begin{table}[t!]
\begin{centering}
\caption{ Comparisons for BEL LASSO/SCAD, soft thresholding,hard thresholding and mean with $\rho=0.3$.}
\label{tb:mean_rho_03} 
\vskip 2mm
\scriptsize{
\begin{tabular}{|c|c||c|c|c||c|c|c||c|c|c||c|c|c||c|}
\hline 
\multirow{2}{*}{$n$}  & \multirow{2}{*}{$p$}  & \multicolumn{3}{c||}{BEL LASSO}  & \multicolumn{3}{c||}{BEL SCAD} & \multicolumn{3}{c||}{Soft thresholding} & \multicolumn{3}{c||}{Hard thresholding} & Mean\tabularnewline
\cline{3-15} 
\multirow{2}{*}{}  & \multirow{2}{*}{}  & MSE  & True  & False & MSE  & True  & False  & MSE  & True  & False  & MSE  & True  & False  & MSE\tabularnewline
\hline 
\hline 
\multirow{2}{*}{$100$}  
& $10$  & $12.2$ & $4.84$ & $0.51$ & $11.8$ & $4.86$ & $0.45$ & $30.8$ & $4.83$ & $0.67$ & $14.3$ & $4.83$ & $0.67$ & $20.5$\tabularnewline
\cline{2-15} 
 & $20$  & $7.9$ & $4.75$ & $0.9$ & $9.8$ & $4.80$ & $1.20$ & $15.1$ & $4.77$ & $2.42$ & $14.3$ & $4.77$ & $2.42$ & $23.1$\tabularnewline
\hline 
\multirow{2}{*}{$200$}  
& $10$  & $6.8$ & $4.85$ & $0.22$  & $8.8$ & $4.81$ & $0.24$ & $22.8$ & $4.83$ & $0.26$ & $7.2$ & $4.83$ & $0.26$ & $9.8$\tabularnewline
\cline{2-15} 
 & $20$  & $3.5$ & $4.77$ & $0.26$ & $4.7$ & $4.79$ & $0.35$ & $13.2$ & $4.77$ & $0.71$ & $5.3$ & $4.77$ & 0.71 & 10.2\tabularnewline
\hline 
\multirow{2}{*}{$500$}  
& $10$  & $2.2$ & $4.97$ & $0.00$ & $2.1$ & $4.97$ & $0.00$ & $22.8$ & $4.97$ & $0.00$ & $2.2$ & $4.97$ & $0.00$ & $3.6$\tabularnewline
\cline{2-15} 
 & $20$  & $1.0$ & $4.97$ & $0.00$ & $1.0$ & $4.99$ & $0.01$ & $11.7$ & $4.96$ & $0.00$ & $1.3$ & $4.96$ & $0.00$ & $3.8$\tabularnewline
\hline 
\end{tabular}}
\par\end{centering}
\end{table}

We compare our methods with the sample mean, the soft
thresholding and the hard thresholding methods. 
The soft thresholding estimates $\mu_j$ using $\hat{\mu}_j^{S} = sign(\bar{\bf X}_{.j})(|\bar{\bf X}_{.j}|-s)_{+}$,
where ${\bf X}_{.j}$ is the $j$th column of matrix ${\bf X}$ and $(c)_+ = c$ for $c>0$ and $0$ otherwise.
The hard thresholding estimates $\mu_j$ using $\hat{\mu}_j^{H} = \bar{\bf X}_{.j} I(|\bar{\bf X}_{.j}|>s)$,
where $I(\cdot)$ is the indicator function.
We choose $s=0.2$ for both the soft thresholding and the hard thresholding. 
The BEL LASSO/SCAD will not set parameters to be exactly equal to $0$.
For comparison, we also use $s$ as the threshold and parameters with the absolute value of the posterior mean less than $s$ will be set to be $0$.
For each combination of $n$ and $p$, $100$ data sets are generated. 
We report the estimated mean squared error (multiplied by $1000$), the number of truly non-zero elements selected and the number of falsely non-zero elements selected in Table \ref{tb:mean_rho_03}  (for $\rho=0.3$) and Table \ref{tb:mean_rho_07}  (for $\rho=0.7$).

From both Table \ref{tb:mean_rho_03} and Table \ref{tb:mean_rho_07}, we can see that the mean squared errors (MSEs) are the smallest for the proposed BEL LASSO/SCAD. Furthermore, the advantages of our methods are much larger for $p=20$.
The proposed methods are able to detect the largest number of the truly non-zero elements. 
At the same time, the numbers of falsely detected non-zero elements are the smallest for our methods. 
When the correlation between each column of $\bf X$ increases, our methods perform the best based on all performance measures. 
In particular, we can see that the advantages of our methods versus other methods increase when we increase the correlation from $0.3$ to $0.7$. 
For example, by comparing Table \ref{tb:mean_rho_03} and Table \ref{tb:mean_rho_07}, we can observe the numbers of falsely detected non-zero elements are about the same for other methods, while those numbers of our methods are greatly reduced from Table \ref{tb:mean_rho_03} to Table \ref{tb:mean_rho_07}. This is because the other methods did not use the correlation information between variables for inference. 
On the other hand, our methods enable us to incorporate that information through the empirical likelihood.

\begin{table}
\begin{centering}
\caption{\label{tb:mean_rho_07}  
Comparisons for BEL LASSO/SCAD, soft thresholding,
hard thresholding and mean with $\rho=0.7$.}
\vskip 2mm
\scriptsize{
\begin{tabular}{|c|c||c|c|c||c|c|c||c|c|c||c|c|c||c|}
\hline 
\multirow{2}{*}{$n$}  & \multirow{2}{*}{$p$}  & \multicolumn{3}{c||}{BEL LASSO}  & \multicolumn{3}{c||}{BEL SCAD} & \multicolumn{3}{c||}{Soft thresholding} & \multicolumn{3}{c||}{Hard thresholding} & Mean\tabularnewline
\cline{3-15} 
\multirow{2}{*}{}  & \multirow{2}{*}{}  & MSE  & True  & False & MSE  & True  & False  & MSE  & True  & False  & MSE  & True  & False  & MSE\tabularnewline
\hline 
\hline 
\multirow{2}{*}{$100$}  & $10$  & $11.0$ & $4.70$ & $0.38$ & $9.9$ & $4.73$ & $0.27$& $29.2$ & $4.68$ & $0.73$ & $15.5$ & $4.68$ & $0.73$ & $18.0$\tabularnewline
\cline{2-15} 
 & $20$  & $2.8$ & $4.82$ & $0.15$ & $3.3$ & $4.84$ & $0.25$ & $14.2$ & $4.77$ & $1.99$ & $12.0$ & $4.77$ & $1.99$ & $17.4$\tabularnewline
\hline 
\multirow{2}{*}{$200$}  & $10$  & $6.1$ & $4.86$ & $0.17$ & $7.9$ & $4.88$ & $0.26$ & $26.6$ & $4.77$ & $0.32$ & $8.9$ & $4.77$ & $0.32$ & $10.0$\tabularnewline
\cline{2-15}
 & $20$  & $1.3$ & $4.95$ & $0.03$ & $1.4$ & $4.93$ & $0.03$& $11.4$ & $4.85$ & $0.74$ & $5.2$ & $4.85$ & $0.74$ & $10.2$\tabularnewline
\hline 
\multirow{2}{*}{$500$}  & $10$  & $1.8$ & $4.99$ & $0.00$ & $2.2$ & $4.97$ & $0.05$ & $20.7$ & $4.97$ & $0.00$ & $2.2$ & $4.97$ & $0.00$ & $4.1$\tabularnewline
\cline{2-15}
 & $20$  & $0.5$ & $4.99$ & $0.00$ & $0.7$ & $4.98$ & $0.00$& $10.4$ & $4.95$ & $0.00$ & $1.1$ & $4.95$ & $0.00$ & $3.8$\tabularnewline
\hline 
\end{tabular}}
\par\end{centering}
\end{table}

\subsection{Simulation for linear regression models}

We simulate the data under the following linear regression model:
$
{\bf y}={\bf X}{\bm \beta} + {\bm \epsilon}, 
$
where ${\bf X} = ({\bf X}_{.1},\ldots,{\bf X}_{.p})$ and ${\bm \beta} = (\beta_1,\ldots,\beta_p)$. 

We consider different combinations of $n$ and $p$, with $n$ being $100$, $200$ or $500$ and $p$ being $10$ or $20$.
We generate $X_{ij}$, $i=1,\ldots,n;j=1,\ldots,p$ from the standard normal distribution.
We conduct simulation studies under the following three scenarios:

\begin{itemize}
\item Scenario A:  The error terms are generated from $\mathcal{N}(0,3^{2})$, and the first five elements of ${\bm \beta}$ are $(1,2,3,4,5)$.
\item Scenario B: The error terms are generated from a mixture of normal distributions, $\mathcal{N}(3,1)$ with probability $0.5$ and $\mathcal{N}(-3,1)$ with probability $0.5$, and the first five elements of ${\bm \beta}$ are $(1,2,3,4,5)$.
\item Scenario C: The error terms are generated from a mixture of normal distributions, $\mathcal{N}(3,1)$ with probability $0.5$ and  $\mathcal{N}(-3,1)$ with probability $0.5$, and the first five elements of ${\bm \beta}$ are $(0.3,0.6,3,4,5)$. Therefore, some of the coefficients are close to $0$.
\end{itemize}

For all the situations, only the first five elements of ${\bm \beta}$ are non-zero. 
For each situation and each combination of $n$ and $p$, we generated $100$ data sets.
We estimate ${\bm \beta}$ using the proposed BEL LASSO/SCAD, Bayesian LASSO (R package "monomvn"), LASSO (R package "glmnet"), SCAD (R package "ncvreg") and ordinary least squares (OLS). 
Both the BEL LASSO/SCAD and Bayesian LASSO require some cut-off values to set parameters to be exactly $0$. 
Therefore, we used 5-fold cross validation to determine the cut-off values. 
For different methods, we compare the MSE (multiplied by $1000$), number of variables correctly identified (True) and incorrectly identified (False). MSE is calculated as the MSE between the estimated ${\bm \beta}$ and its true value. Since for each situation, there are $5$ variables in the model, we expect "True" to be close to $5$ and "False" to be close to $0$.

\begin{table}
\begin{centering}
\caption{\label{tb:table3}
Comparisons for BEL LASSO/SCAD, Bayesian LASSO, LASSO, SCAD and
OLS when the error terms follow normal distributions.} 
\vskip 2mm
\scriptsize{
\begin{tabular}{|c||c|c||c|c||c|c||c|c||c|c||c|}
\hline 
{\multirow{2}{*}{$n/p$}}  & \multicolumn{2}{c||}{BEL LASSO}  & \multicolumn{2}{c||}{BEL SCAD} & \multicolumn{2}{c||}{Bayesian LASSO} & \multicolumn{2}{c||}{LASSO} & \multicolumn{2}{c||}{SCAD} & OLS \tabularnewline
\cline{2-12} 
{\multirow{2}{*}{}} & M  & T/F/F$_1$  & M  & T/F/F$_1$ & M  & T/F/F$_1$ & M  & T/F/F$_1$ & M  & T/F/F$_1$  & M\tabularnewline
\hline 
\hline 
 $100/10$  & $87$  & $4.9/1.6/88$ & ${\bf 84}$  & $5.0/1.3/89$   & $85$  & $5.0/1.3/{\bf 90}$ & $98$  & $5.0/3.1/77$ & $90 $  & $4.9/1.2/90 $ & $103$\tabularnewline
\hline 
 $100/20$  & $67$  & $4.9/2.0/84$  & $64$  & $4.9/1.9/85$ & $ 54 $  & $4.9/1.0/{\bf 90} $ & $67$ & $5.0/5.6/66 $ & ${\bf 46} $  & $ 4.9/1.7/86 $ & $110$\tabularnewline
\hline 
 $200/10$  & $35$  & $5.0/0.8/94$  & ${\bf 33}$  & $5.0/0.6/{\bf 96}$  & $36 $  & $5.0/0.8/94 $ & $ 42$  & $5.0/2.9/79 $ & $38$& $5.0/1.3/90$ & $47$\tabularnewline
\hline 
 $200/20$  & $21$  & $5.0/0.9/93$  & $20$  & $5.0/0.9/94$  & $19$  & $5.0/0.7/{\bf 95}$ & $ 30$  & $5.0/5.8/66 $ & ${\bf 18}$ & $5.0/1.8/87 $ & $49$\tabularnewline
\hline 
 $500/10$  & ${\bf 13}$  & $5.0/0.4/{\bf 97}$  & ${\bf 13}$  & $5.0/0.4/{\bf 97}$  & ${\bf 13} $  & $5.0/0.3/{\bf 97} $ & $18 $  & $5.0/3.0/78 $ & $15 $ & $5.0/0.8/94$ & $19$\tabularnewline
\hline 
 $500/20$  & ${\bf 6}$  & $5.0/0.3/98$  & ${\bf 6}$  & $5.0/0.1/{\bf 99}$  & $ {\bf 6}$  & $5.0/0.3/98 $ & $11 $  & $ 5.0/5.3/68$ & $7$ & $5.0/0.8/94 $ & $18$\tabularnewline
\hline 
\end{tabular}}
\par\end{centering}
\end{table}

\begin{table}
\begin{centering}
\caption{\label{tb:table4} 
Comparisons for BEL LASSO/SCAD, Bayesian LASSO, LASSO, SCAD and
OLS when the error terms follow a mixture of normal distributions.}
\vskip 2mm
\scriptsize{
\begin{tabular}{|c||c|c||c|c||c|c||c|c||c|c||c|}
\hline 
{\multirow{2}{*}{$n/p$}}  & \multicolumn{2}{c||}{BEL LASSO}  & \multicolumn{2}{c||}{BEL SCAD} & \multicolumn{2}{c||}{Bayesian LASSO} & \multicolumn{2}{c||}{LASSO} & \multicolumn{2}{c||}{SCAD} & OLS \tabularnewline
\cline{2-12} 
{\multirow{2}{*}{}} & M  & T/F/F$_1$  & M  & T/F/F$_1$ & M  & T/F/F$_1$ & M  & T/F/F$_1$ & M  & T/F/F$_1$  & M\tabularnewline
\hline 
\hline 
 $100/10$  & ${\bf 72}$  & $5.0/1.0/{\bf 92}$ & $73$  & $5.0/1.3/90$   & $88$  & $5.0/1.2/90$ & $98$  & $5.0/2.9/78$ & $93 $  & $4.9/1.2/90 $ & $110$\tabularnewline
\hline 
 $100/20$  & $64$  & $4.9/2.0/84$  & $58$  & $4.9/1.8/{\bf 86}$ & $ 71 $  & $4.9/2.4/83 $ & $75$ & $5.0/6.2/64 $ & ${\bf 54} $  & $ 4.9/2.4/82 $ & $130$\tabularnewline
\hline 
 $200/10$  & $37$  & $5.0/0.6/95$  & ${\bf 30}$  & $5.0/0.5/{\bf 96}$  & $34 $  & $5.0/0.5/{\bf 96} $ & $ 44$  & $5.0/2.9/78 $ & $39$& $5.0/1.2/91$ & $51$\tabularnewline
\hline 
 $200/20$  & $20$  & $5.0/0.8/{\bf 94}$  & ${\bf 19}$  & $5.0/0.8/{\bf 94}$  & $23$  & $5.0/0.7/{\bf 94} $ & $ 33$  & $5.0/5.8/65 $ & $23$ & $5.0/1.9/85 $ & $55$\tabularnewline
\hline 
 $500/10$  & ${\bf 10}$  & $5.0/0.4/{\bf 97}$  & $11$  & $5.0/0.5/{\bf 97}$  & $11 $  & $5.0/0.5/{\bf 97} $ & $16 $  & $5.0/2.9/79 $ & $11 $ & $5.0/0.6/95$ & $18$\tabularnewline
\hline 
 $500/20$  & ${\bf 7}$  & $5.0/0.3/{\bf 98}$  & ${\bf 7}$  & $5.0/0.8/96$  & $ 8$  & $5.0/0.7/95 $ & $14 $  & $ 5.0/6.7/62$ & $8$ & $5.0/1.4/90 $ & $22$\tabularnewline
\hline 
\end{tabular}}
\par\end{centering}
\end{table}

\begin{table}
\begin{centering}
\caption{\label{tb:table42} 
Comparisons for BEL LASSO/SCAD, Bayesian LASSO, LASSO, SCAD and
OLS when the error terms follow a mixture of normal distributions and some of the coefficients are close to $0$.}
\vskip 2mm
\scriptsize{
\begin{tabular}{|c||c|c||c|c||c|c||c|c||c|c||c|}
\hline 
{\multirow{2}{*}{$n/p$}}  & \multicolumn{2}{c||}{BEL LASSO}  & \multicolumn{2}{c||}{BEL SCAD} & \multicolumn{2}{c||}{Bayesian LASSO} & \multicolumn{2}{c||}{LASSO} & \multicolumn{2}{c||}{SCAD} & OLS \tabularnewline
\cline{2-12} 
{\multirow{2}{*}{}} & M  & T/F/F$_1$  & M  & T/F/F$_1$ & M  & T/F/F$_1$ & M  & T/F/F$_1$ & M  & T/F/F$_1$  & M\tabularnewline
\hline 
\hline 
 $100/10$  & ${\bf 74}$  & $4.1/1.1/{\bf 81}$ & $75$  & $4.1/1.1/{\bf 81}$   & $86$  & $4.1/1.2/80$ & $94$  & $4.5/2.5/76$ & $80$  & $3.8/0.8/80 $ & $111$\tabularnewline
\hline 
 $100/20$  & $60$  & $3.6/1.4/73$  & $62$  & $3.7/2.1/71$ & $ 67 $  & $3.7/1.8/72 $ & $70$ & $4.2/5.1/61 $ & ${\bf 43} $  & $ 3.5/1.2/{\bf 74} $ & $131$\tabularnewline
\hline 
 $200/10$  & $37$  & $4.1/0.7/84$  & ${\bf 35}$  & $4.3/0.7/{\bf 86}$  & $41 $  & $4.2/0.9/83 $ & $ 41$  & $4.7/2.7/77 $ & $43$& $4.1/1.0/82$ & $47$\tabularnewline
\hline 
 $200/20$  & $25$  & $3.9/1.0/81$  & ${\bf 24}$  & $4.0/0.8/{\bf 83}$  & $27$  & $4.0/1.0/81 $ & $ 32$  & $4.6/5.3/64 $ & $25$ & $4.0/1.3/79 $ & $56$\tabularnewline
\hline 
 $500/10$  & $18$  & $4.5/0.8/88$  & ${\bf 17}$  & $4.5/0.8/{\bf 89}$  & $18 $  & $4.5/0.7/{\bf 89} $ & $19 $  & $4.9/3.1/76 $ & $21 $ & $4.7/1.5/85$ & $21$\tabularnewline
\hline 
 $500/20$  & $10$  & $4.4/0.8/88$  & ${\bf 9}$  & $4.3/0.6/88$  & $ 10$  & $4.3/0.5/{\bf 89} $ & $12 $  & $ 4.9/5.9/65$ & $10$ & $4.6/2.2/81 $ & $21$\tabularnewline
\hline 
\end{tabular}}
\par\end{centering}
\end{table}

The averaged MSEs (multiplied by $1000$) of ${\bm \beta}$ (column ``M'') along with the numbers of truly/falsely discovered covariates and the F$_1$ (multiplied by $100$) scores (column ``T/F/F$_1$'') are reported in Tables \ref{tb:table3} -- \ref{tb:table42} for situations A -- C, respectively. The F$_1$ score is defined as the harmonic mean of precision and recall: 
\begin{equation*}
F_1 = \left(\frac{2}{recall^{-1}+precision^{-1}}\right) = 2 \frac{precision\cdot recall}{precision + recall},
\end{equation*}
where precision is defined as the ratio between the number of true positives (T) over the number of total positives (T+F); 
recall is defined as the ratio between the number of true positives (T) over the number of actual positives ($5$ in our examples).
For each setup, we highlight the smallest MSE and the largest F$_1$ in bold.
 In general, we can see that the performance of different methods is better for the larger sample size. 
 The MSE decreases as the sample size increases, and the F$_1$ score increases as the sample size increases.
 The MSEs for all shrinkage based methods (BEL LASSO, BEL SCAD, Bayesian LASSO, LASSO, SCAD) are smaller than those of OLS. 
 Also, the LASSO always tends to select a large number of extra covariates, which are not in the true model.

When the error terms are normally distributed, the performance of different methods are comparable in terms of MSE and number of correctly selected covariates. 
LASSO tends to select most number of false covariates, followed by the SCAD.  
When the error terms follow a mixture of two normal distributions and the non-zero elements of ${\bm \beta}$ are $(1,2,3,4,5)$, 
the proposed methods produce the smallest MSEs and the largest F$_1$ scores in almost all of the cases. 
Among the shrinkage based methods, the LASSO tends to produce the largest MSE and select the largest number of unnecessary covariates. 
All methods are able to identify almost all of the true covariates.   

When the non-zero elements of ${\bm \beta}$ are large enough in magnitude, all the methods are able to identify the true covariates. 
Thus, in Scenario C, we set two of the coefficients to be small ($0.3$ and $0.6$) and compare the simulation results. 
As we can see in Table \ref{tb:table42}, when the signals become weaker, the number of true covariates being selected becomes less. 
In general, the LASSO selects the most number of true covariates. 
However, it also selects more  fasle covariates. 
Also, the LASSO produces the largest MSEs and the smallest F$_1$ scores among all the shrinkage based methods. 
It is interesting to see that SCAD performs the best among all the methods when $n=100$ and $p=20$. 
However, 
as the sample size increases, both SCAD and LASSO tend to select more false covariates, while the trend is opposite for the Bayesian methods. 
Our methods produce the smallest MSE, among all the methods in most of the cases.
When the sample size is smaller ($n=100$ or $n=200$), our methods produce the largest F$_1$ scores in most cases, while the Bayesian LASSO wins for $n=500$.

{ \section{Real data analysis}}

\subsection{Ozone data}

We use the well-known ozone data from Breiman and Friedman (1985). The data was collected in 1976 in the area of Upland, CA, east of Los Angeles. 
In addition to the 1-hour average ozone level, we have information on time related variables including month, day, and enviorment related variables including
500-millibarpressure height (P500H), wind speed, humidity, temperature at Sandberg (tempS), temperature at El Monte (tempE), inversion base height (IBH),
pressure gradient from Los Angeles International Airport to Daggett (PGD), inversion base temperature (IBT) and square root of visibility. 
The data set is available in R packages "bfp".
We are interested in knowing how these information are related to the ozone level. All the variables including the response are standardized to have mean $0$ and standard deviation $1$. We report the $95\%$ confidence intervals
and the corresponding interval lengths for each coefficient obtained using BEL LASSO, BEL SCAD, Bayesian LASSO and OLS. 
The results are reported in Table \ref{tb:table5b}.
We do not include the results for LASSO and SCAD, since they do not produce confidence intervals for the estimates.

\begin{table}
\begin{centering}
\caption{\label{tb:table5b}
$95\%$ confidence intervals and interval lengths for the parameters estimated using BEL LASSO/SCAD, Bayesian LASSO and OLS using the ozone data.}
\vskip 2mm
\scriptsize{
\begin{tabular}{|c|c|c|c|c|c|c|c|c|}
\hline 
\multirow{2}{*}{ }  & \multicolumn{2}{c|}{BEL LASSO} & \multicolumn{2}{c|}{BEL SCAD} & \multicolumn{2}{c|}{Bayesian LASSO} & \multicolumn{2}{c|}{OLS}\tabularnewline
\cline{2-9}
& C.I. & Len. & C.I. & Len. & C.I. & Len. &C.I. & Len.\\
\hline 
day  & $(-0.05,0.07)$ & $0.12$ & $(-0.06,0.10)$ & $0.16$ & $(-0.06,0.08)$ & $0.14$ & $(-0.06,0.10)$ & $0.16$\\
\hline 
P500H & $(-0.26,0.02)$& $0.28$ & $(-0.33,0.03)$& $0.36$ & $(-0.26,0.07)$ & $0.33$  & $(-0.39,0.00)$& $0.39$\\
\hline 
WindSpeed  & $(-0.10,0.06)$ & $0.16$ & $(-0.07,0.09)$ & $0.16$  & $(-0.06,0.10)$ & $0.16$  & $(-0.08,0.09)$ & $0.17$\\
\hline 
humidity & $(0.09,0.30)$ & $0.21$ & $(0.10,0.33)$ & $0.23$ & $(0.10,0.31)$ & $0.21$  & $(0.09,0.33)$ & $0.24$ \\
\hline 
tempS & $(0.19,0.40)$ & $0.21$ & $(0.01,0.41)$ & $0.40$ & $(0.01,0.43)$ & $0.42$ & $(0.01,0.49)$ & $0.48$\\
\hline 
tempE  & $(0.55,0.76)$ & $0.21$ & $(0.33,0.93)$ & $0.60$& $(0.17,0.78)$ & $0.61$ & $(0.41,1.12)$ & $0.71$\\
\hline 
IBH  & $(-0.33,-0.15)$ & $0.18$& $(-0.30,-0.03)$ & $0.27$ & $(-0.24,0.00)$ & $0.24$ & $(-0.40,-0.05)$ & $0.35$\\
\hline 
PGD & $(-0.05,0.13)$ & $0.18$ & $(-0.07,0.18)$ & $0.25$& $(-0.04,0.17)$ & $0.21$ & $(-0.10,0.16)$ & $0.26$\\
\hline 
IBT & $(-0.41,-0.13)$ & $0.28$ & $(-0.37,0.23)$ & $0.60$& $(-0.21,0.27)$ & $0.48$ & $(-0.67,0.14)$ & $0.81$\\
\hline 
visibility & $(-0.09,0.06)$ & $0.15$ & $(-0.11,0.04)$ & $0.15$& $(-0.13,0.05)$ & $0.18$ & $(-0.13,0.01)$ & $0.14$\\
\hline 
\end{tabular}}
\par\end{centering}
\end{table}
 
We observe that, in general, the variable selection based methods produced confidence intervals with shorter widths than OLS. 
Amongth the variable selection based methods, BEL LASSO always generate the intervals with the shortest widths, except for visibility.
BEL LASSO identified humidity, temperature at Sandberg, temperature at El Monte, inversion base height and inversion base temperature as the important variables. 
BEL SCAD and Bayesian LASSO find similar results except they did not include the inversion base temperature.

\subsection{Wage data}
The wage data contains information on workers' attributes that might affect the hourly wage. 
The dependent variable is the logrithm of the hourly wage. 
The predictor variables include education in years, potential experience in years, ability, mother's education in years (MEdu),
father's education in years (FEdu), broken home until age of 14, and number of siblings. 
This data set has been analyzed by Koop and Tobias (2004) and is shared in the Journal of Applied Econometrics data archive. 
For our analysis, we standardized all the variables to have mean $0$ and standard deviation $1$.

The $95\%$ confidence intervals and interval lengths are calculated using BEL LASSO, BEL SCAD, Bayesian LASSO and OLS. 
The results are reported in Table \ref{tb:table5a}. We observe that the confidence intervals obtained using different methods are similar with most of the variables identified to be important except mother's education in years and number of siblings. 
It is interesting that father's education is found important while mother's education not.

\begin{table}
\begin{centering}
\caption{\label{tb:table5a}
$95\%$ confidence intervals and interval lengths for the parameters estimated using BEL LASSO/SCAD, Bayesian LASSO and OLS using the wage data.}
\vskip 2mm
\scriptsize{
\begin{tabular}{|c|c|c|c|c|c|c|c|c|}
\hline 
\multirow{2}{*}{ }  & \multicolumn{2}{c|}{BEL LASSO} & \multicolumn{2}{c|}{BEL SCAD} & \multicolumn{2}{c|}{Bayesian LASSO} & \multicolumn{2}{c|}{OLS}\tabularnewline
\cline{2-9}
& C.I. & Len. & C.I. & Len. & C.I. & Len. &C.I. & Len.\\
\hline 
Edu  & $(0.29,0.33)$ & $0.04$ & $(0.29,0.33)$ & $0.04$ & $(0.29,0.33)$ & $0.04$ & $(0.29,0.33)$ & $0.04$\\
\hline 
Exp & $(0.41,0.45)$& $0.04$ & $(0.40,0.45)$& $0.05$ & $(0.40,0.45)$ & $0.05$  & $(0.40,0.45)$& $0.05$\\
\hline 
Time  & $(-0.16,-0.11)$ & $0.05$ & $(-0.16,-0.11)$ & $0.05$  & $(-0.16,-0.11)$ & $0.05$  & $(-0.16,-0.11)$ & $0.05$\\
\hline 
Abi & $(0.12,0.16)$ & $0.04$ & $(0.12,0.16)$ & $0.04$ & $(0.12,0.16)$ & $0.04$  & $(0.12,0.16)$ & $0.04$ \\
\hline 
MEdu  & $(-0.03,0.01)$ & $0.04$ & $(-0.02,0.02)$ & $0.04$ & $(-0.02,0.01)$ & $0.03$ & $(-0.02,0.01)$ & $0.03$\\
\hline 
FEdu  & $(0.02,0.07)$ & $0.05$ & $(0.02,0.06)$ & $0.04$& $(0.02,0.06)$ & $0.04$ & $(0.02,0.060)$ & $0.04$\\
\hline 
BH  & $(-0.05,-0.02)$ & $0.03$& $(-0.05,-0.02)$ & $0.03$ & $(-0.04,-0.02)$ & $0.02$ & $(-0.05,-0.02)$ & $0.03$\\
\hline 
Sib & $(-0.00,0.03)$ & $0.03$ & $(-0.00,0.03)$ & $0.03$& $(-0.00,0.03)$ & $0.03$ & $(-0.00,0.03)$ & $0.03$\\
\hline 
\end{tabular}}
\par\end{centering}
\end{table}

\subsection{Prostate cancer data}
We use the same prostate cancer data (Stamery et al. 1989) as used in Tibshirani (1996). 
The dependent variable is the $\log$ of the level of prostate specific antigen (lpsa).
The predictors include $8$ clinical measures: $\log($cancer volumn$)$ (lcaval), $\log($prostate weight$)$ (lweight), age, $\log($benign prostatic hyperplasia amount$)$ (llbph), seminal vesicle invasion (svi), $\log($capsular penetration$)$ (lcp), Gleason score (gleason)  and percentage Gleason scores $4$ or $5$ (pgg45). The data set is available in R package ``lasso2''.

To perform the analysis, we standardize the predictors such that they all have mean $0$ and standard deviation $1$.
The dependent variable is also standardized to have mean $0$.
We report the $95\%$ confidence intervals in Table \ref{tb:table5} for each coefficient obtained using BEL LASSO/SCAD, Bayesian LASSO and OLS and the corresponding interval lengths.

\begin{table}
\begin{centering}
\caption{\label{tb:table5}
$95\%$ confidence intervals and interval lengths for the parameters estimated using BEL LASSO/SCAD, Bayesian LASSO and OLS using the prostate cancer data.}
\vskip 2mm
\scriptsize{
\begin{tabular}{|c|c|c|c|c|c|c|c|c|}
\hline 
\multirow{2}{*}{ }  & \multicolumn{2}{c|}{BEL LASSO} & \multicolumn{2}{c|}{BEL SCAD} & \multicolumn{2}{c|}{Bayesian LASSO} & \multicolumn{2}{c|}{OLS}\tabularnewline
\cline{2-9}
& C.I. & Len. & C.I. & Len. & C.I. & Len. &C.I. & Len.\\
\hline 
lcavol  & $(0.53,0.82)$ & $0.29$ & $(0.48,0.85)$ & $0.37$ & $(0.37,0.76)$ & $0.39$ & $(0.49,0.89)$ & $0.40$\\
\hline 
lweight & $(0.10,0.38)$& $0.28$ & $(0.08,0.43)$& $0.35$ & $(0.06,0.41)$ & $0.35$  & $(0.06,0.39)$& $0.33$\\
\hline 
age  & $(-0.26,-0.02)$ & $0.24$ & $(-0.28,0.01)$ & $0.29$  & $(-0.23,0.07)$ & $0.30$  & $(-0.31,0.02)$ & $0.33$\\
\hline 
lbph & $(0.01,0.26)$ & $0.25$ & $(-0.03,0.29)$ & $0.32$ & $(-0.05,0.25)$ & $0.30$  & $(-0.01,0.32)$ & $0.33$ \\
\hline 
svi  & $(0.17,0.45)$ & $0.28$ & $(0.13,0.50)$ & $0.37$ & $(0.04,0.43)$ & $0.39$ & $(0.12,0.51)$ & $0.39$\\
\hline 
lcp  & $(-0.31,0.04)$ & $0.35$ & $(-0.34,0.11)$ & $0.45$& $(-0.20,0.16)$ & $0.36$ & $(-0.40,0.10)$ & $0.50$\\
\hline 
lgleason  & $(-0.12,0.18)$ & $0.30$& $(-0.18,0.22)$ & $0.40$ & $(-0.11,0.20)$ & $0.31$ & $(-0.19,0.25)$ & $0.44$\\
\hline 
pgg45 & $(-0.06,0.30)$ & $0.36$ & $(-0.11,0.36)$ & $0.47$& $(-0.08,0.26)$ & $0.34$ & $(-0.12,0.37)$ & $0.49$\\
\hline 
\end{tabular}}
\par\end{centering}
\end{table}

From the results, we observe that all of the methods identified three important variables that should be kept in the model: $\log($cancer volumn$)$ (lcaval), $\log($prostate weight$)$ (lweight),  and seminal vesicle invasion (svi). There are two other variables with the boundaries of the confidence intervals close to $0$: age and $\log($benign prostatic hyperplasia amount$)$ (llbph). The remianing three variables are not selected by any of the methods.

\begin{figure}[htbp] 
\begin{center}
\begin{tabular}{c}
\epsfig{figure=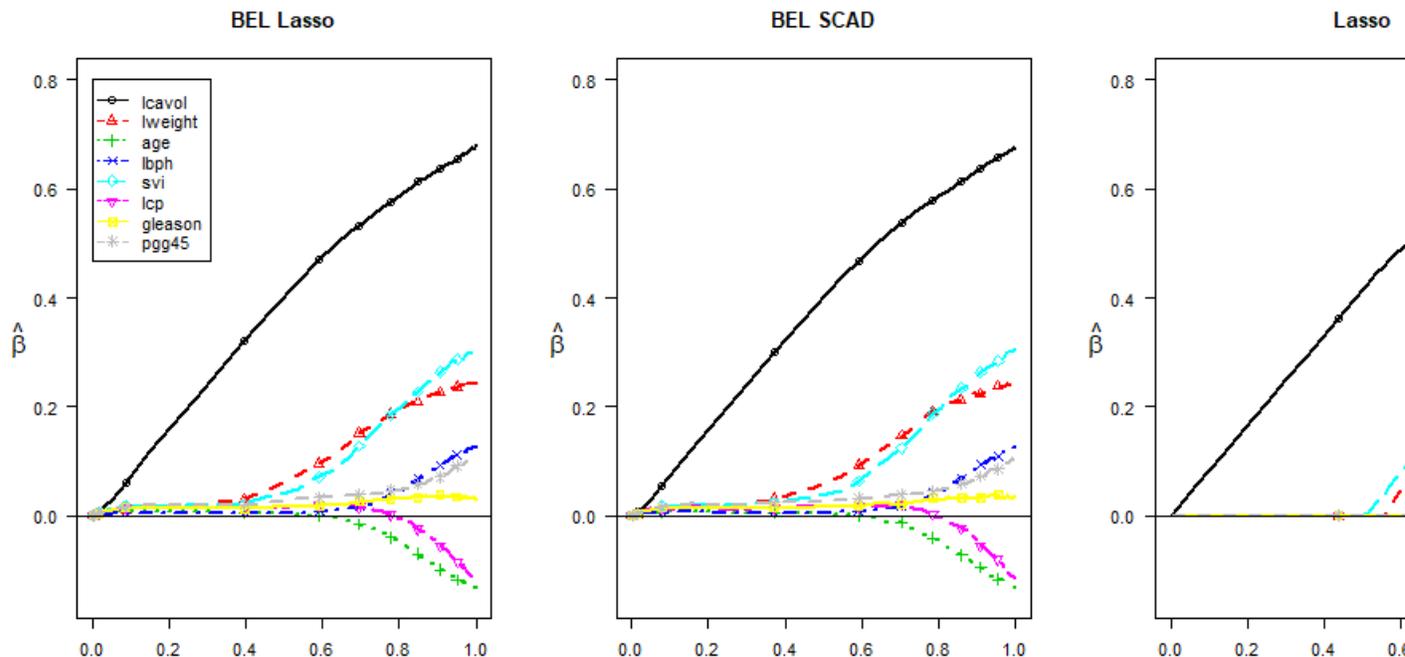,scale = .6} 
\end{tabular}
\caption{ The solution path of $\bm \hat{\beta}$ for BEL LASSO, BEL SCAD and LASSO. The x axis is the ratio between the $L_2$ norm of the estimate and the maximum $L_2$ norm.}
\label{fig:solution_path}
\end{center}
\end{figure}

For comparison purpose, we also include the solution paths for our methods and the Lasso in Figure \ref{fig:solution_path} , the $x$ axis is the ratio between the $L_2$ norm of the estimate and the maximum $L_2$ norm. It is interesting to observe that, overall, the solution paths for the three methods are similar. However, the two versions of our method do not set the estimate to be exactly zero but some small value close to zero.

{ \section{Discussion}}
In this paper, we proposed a novel semiparametric variable selection method based on empirical likelihood. 
We showed that a general class of prior distributions leads to both consistent posteriors and variable selection consistency as long as the prior assigns non-trivial mass around the truth. 
As special cases, we show that the Laplace prior and the SCAD prior belong to this general class. 
In addition, it is easy to verify that this class also includes the ridge prior and the elastic net prior, which establishes the 
close connection between our method and those of Bedoui and Lazar (2020) and Moon and Bedoui (2020).
The numerical results show that the proposed method can outperform its parametric counterparts, especially when the model for the error term is misspecified. 

Sampling from the posterior distribution for the empirical likelihood based method can be challenging. 
This is due to the fact that the Lagrange multiplier ${\bm \lambda}$ in an empirical likelihood needs to be solved numerically. 
As a result, there exists no conjugate prior for the empirical likelihood and Metropolis-Hasting type of sampling methods need to be applied for the sampling of a posterior distribution. 
This problem is enlarged especially for high-dimensional situations.  
We make use of the fact that the empirical likelihood can be approximated by a multivariate normal distribution and a novel approximation based sampling method is  proposed. 
The computational results demonstrate the quick convergence. 

Limited by the nature of the empirical likelihood, the proposed method only works when $p$ grows at a slower rate than $n$. This is due to the fact that the equations of ${\bm \lambda}$ have no solution when $p = O(n^h)$ for $h\geq1/2$. 
To overcome this limitation, several recent EL methods are proposed, including Chen et al. (2008); Lahiri and Mukhopadhyay (2012); and Chang et al. (2018), among others. 
Their work provides the opportunity to extend our proposed procedure to the ultrahigh dimensional setting. We leave this for future endeavors.

%%%%%%%%%%%%%%%%%%%%%%%%%%%%%%%%%%%%%%%%%%%%%%%%%%%%%%%%%%%%%%%%%%%%%%%%%%%%%%%%%%%%%%%%%%%%%%%%%%%%%%%%%%%%%%%%%%%%%%%%%%%%

{\centering \section*{SUPPLEMENTARY MATERIAL}}

The online supplementary materials contain further techinical details.
%%%%%%%%%%%%%%%%%%%%%%%%%%%%%%%%%%%%%%%%%%%%%%%%%%%%%%%%%%%%%%%%%%%%%%%%%%%%%%%%%%%%%%%%%%%%%%%%%%%%%%%%%%%%%%%%%%%%%%%%%%%%

\vspace{1cm}

{ \section*{References} }
%%%%%%%%%%%%%%%%%%%%%%%%%%%%%%%%%%%%%%%%%%%%%%%%%%%%%%%%%%
\begin{description}

\item[] Armagan, A., D.B. Dunson, J. Lee, W.U. Bajwa, and N. Strawn (2013). ``Posterior Consistency in Linear Models under Shrinkage
Priors." \textit{Biometrika} {\it 100}(4), 1011\textendash 1018.

\item[] Bartolucci, F. (2007). ``A Penalized Version of the Empirical Likelihood Ratio for the Population Mean."
\textit{Statistics \& Probability Letters} {\it 77}(1), 104--110.

\item[] Bhattacharya, A., D. Pati, N.S. Pillai and D.B. Dunson (2015).
``Dirichlet--Laplace Priors for Optimal Shrinkage."
\textit{ Journal of the American Statistical Association} {\it 110}(512), pp. 1479--1490.

\item[] Bedoui, A., and N. Lazar N. (2020). ``Bayesian Empirical Likelihood for Ridge and Lasso Regressions."
\textit{ Computational Statistics and Data Analysis} {\it 145}, in press.

\item[] Breiman, L., and J.H. Friedman (1985). ``Estimating Optimal Transformations for Multiple Regression and Correlation."
\textit{ Journal of the American Statistical Association} {\it 80}(391), 580--598.

\item[] Casella, G. (2001). ``Empirical Bayes Gibbs Sampling." \textit{Biostatistics} {\it 2}(4), 485--500.

\item[] Castillo, I., J. Schmidt-Hieber, and A. van der Vaart (2015). ``Bayesian Linear Regression With Sparse Priors."
\textit{Annals of Statistics} {\it 43}(5), 1986--2018.

\item[]  Chang J., S.X. Chen, and X. Chen (2015). ``High Dimensional Generalized Empirical Likelihood for Moment Restrictions with Dependent Data." 
\textit{Journal of Econometrics} {\it 185}(1), 283--304. 

\item[]  Chang J., C.Y. Tang, and T.T.  Wu (2018). ``A New Scope of Penalized Empirical Likelihood with High-dimensional Estimating Equations." 
\textit{The Annals of Statistics} {\it 46}(6B), 3185--3216. 

\item[]  Chen J., A.M. Variyath, and B. Abraham (2008). ``Adjusted Empirical Likelihood and Its Properties." 
\textit{Journal of Computational and Graphical Statistics} {\it 17}(2), 426--443. 

\item[]  Chen, S.X. (1996). ``Empirical Likelihood Confidence Intervals for Nonparametric Density Estimation." \textit{Biometrika} {\it 83}(2),
329\textendash 341. 

\item[]  Chen, S.X., L. Peng, and Y.L. Qin (2009). ``Effect of Data Dimension on Empirical Likelihood." \textit{Biometrika} {\it 196}(3), 711--722.

\item[]  Chen, S.X., and I. Van Keilegom (2009). ``A Review on Empirical Likelihood Methods for Regression." \textit{Test} {\it 18},
415\textendash 447. 

\item[] Chib S., M. Shin, and A. Simoni (2018). ``Bayesian Estimation and Comparison of Moment Condition Models." \textit{Journal of the American Statistical Association} 
{\it 113}(524), 1656--1668.

\item[]  Chrzanowski, M. (2014). ``Weighted Empirical Likelihood Inference for the Area under the ROC Curve." 
\textit{Journal of Statistical Planning and Inference} {\it 147}, 159--172.

\item[]  Diciccio, T.J., P. Hall, and J.P. Romano (1991). ``Empirical Likelihood is Bartlett Correctable."
\textit{Annals of Statistics} {\it 19}(2), 1053\textendash 1061. 

\item[]  Efron, B., T. Hastie, I. Johnstone, and R. Tibshirani (2004). ``Least Angle Regression."
\textit{Annals of Statistics} {\it 32}(2), 407\textendash 499. 

\item[]  Fan, J., and  R. Li (2001). ``Variable Selection via Nonconcave Penalized Likelihood and its Oracle Properties."
\textit{Journal of the American Statistical Association} {\it 96}(456), 1348\textendash 1360. 

\item[]  Fan, J., and H. Peng (2004). ``Nonconcave Penalized Likelihood with a Diverging Number of Parameters."
\textit{The Annals of Statisitcs} {\it 32}(3), 928\textendash 961. 

\item[]  Hall, P. and A.B. Owen (1993). ``Empirical Likelihood Confidence Bands in Density Estimation."
\textit{Journal of Computational and Graphical Statistics} {\it 2}(2), 273\textendash289.

\item[] Hjort, H.L., I.W. McKeague, and I. Van Keilegom (2008). ``Extending the Scope of Empirical Likelihood.''
\textit{The Annals of Statisitcs} {\it 37}(3), 1079\textendash1115.

\item[] Jing, B.Y., J. Yuan, and Z. Wang (2009). ``Jackknife Empirical Likelihood."
\textit{Journal of the American Statistical Association} {\it 104}(487), 319--326.

\item[] Koop, G., and J.L. Tobias (2004). ``Learning about Heterogeneity in Returns to Schooling."
\textit{Journal of Applied Econometrics} {\it 19}(7), 827--849.

\item[]  Lahiri, S.N., and S. Mukhopadhyay (2012). ``A Penalized Empirical Likelihood Method in High Dimensions."
\textit{The Annals of Statisitcs} {\it 40}(5), 2511\textendash2540.

\item[]  Lancaster, T., and S.J. Jun (2010). ``Bayesian Quantile Regression Methods.'' \textit{Journal of Applied Econometrics}
{\it 25}(2), 287--307.

\item[]  Lazar, N.A. (2003). ``Bayesian Empirical Likelihood." \textit{Biometrika} {\it 90}(2), 319\textendash326.

\item[]  Leng, C., and C.Y. Tang  (2012). ``Penalized Empirical Likelihood and Growing Dimensional General Estimating Equations."
\textit{Biometrika} {\it 99}(3), 703\textendash716.

\item[] Li, J. (2011). ``The Bayesian Lasso, Bayesian Scad and Bayesian Group Lasso With Applications to Genome-Wide Association Studies."
\textit{PhD Thesis, The Pennsylvania State University}.

\item[]  Monahan, J.F., and D.D. Boos (1992). ``Proper Likelihoods for Bayesian Analysis.'' \textit{Biometrika}  {\it 79}(2), 271--278.

\item[] Moon, C. and A. Bedoui (2020). ``Bayesian Elastic Net based on Empirical Likelihood." \textit{arXiv:2006.10258}.

\item[]  Owen, A.B. (1988). ``Empirical Likelihood Ratio Confidence Intervals for a Single Functional." \textit{Biometrika} {\it 75}(2), 237--249.

\item[]  Owen, A.B. (1990). ``Empirical Likelihood Confidence Regions." \textit{Annals of Statistics} {\it 18}(1), 90--120.

\item[] Park, T., and G. Casella (2008). ``The Bayesian Lasso." \textit{Journal of the American Statistical Association} {\it 103}(482),
681\textendash 686.

\item[] Qin, G., and X. Zhou (2006). ``Empirical Likelihood Inference for the Area under the ROC Curve." \textit{Biometrics} {\it 62}(2): 613–622. 

\item[] Schennach, S.M. (2005). ``Bayesian Exponentially Tilted Empirical Likelihood.'' \textit{Biometrika}  {\it 92}(1): 31--46.

\item[] Stamery, T., J. Kabalin, J. McNeal, I. Johnstone, G. Freiha, E. Redwine, and N. Yang  (1989). ``Prostate Specifice Antigen in the Diagnosis and Treatment of Adenocarcinoma of the Prostate, II: Radical Postatectomy Treated Patients."
\textit{Journal of Urology} {\it 141}(5), 1076\textendash 1083.

\item[] Tang, C.Y., and C. Leng (2010). ``Penalized High-dimensional Empirical Likelihood." \textit{Biometrika} {\it 97}(4), 905\textendash 920.

\item[] Tibshirani, R. (1996). ``Regression Shrinkage and Selection via the Lasso." \textit{Journal of the Royal Statistical Society, Ser. B} {\it 58}(1), 267\textendash 288.

\item[] Yang, H., and Y. Zhao (2013). ``Smoothed Jackknife Empirical Likelihood Inference for the Difference of ROC Curves." \textit{Journal of Multivariate Analysis} {\it 115}, 270\textendash 284.

\item[] Yang, H., and Y. Zhao (2015). ``Smoothed Jackknife Empirical Likelihood Inference for ROC Curves with Missing Data." \textit{Journal of Multivariate Analysis} {\bf 140}, 123\textendash 138.

\item[] Yang, Y., and X. He (2012). ``Bayesian Empirical Likelihood for Quantile Regression.'' \textit{The Annals of Statistics} {\it 40}(2), 1102--1113.

\end{description}

%%%%%%%%%%%%%%%%%%%%%%%%%%%%%%%%%%%%%%%%%%%%%%%%%%%%%%%%%%%%%%%%%%%%%%%%%%%%%%%%%%%%%%%%%%%%%%%%%%%%%%%%%%%%%%%%%%%%%%%%%%%%
%%%%%%%%%%%%%%%%%%%%%%%%%%%%%%%%%%%%%%%%%%%%%%%%%%%%%%%%%%%%%%%%%%%%%%%%%%%%%%%%%%%%%%%%%%%%%%%%%%%%%%%%%%%%%%%%%%%%%%%%%%%%
%%%%%%%%%%%%%%%%%%%%%%%%%%%%%%%%%%%%%%%%%%%%%%%%%%%%%%%%%%%%%%%%%%%%%%%%%%%%%%%%%%%%%%%%%%%%%%%%%%%%%%%%%%%%%%%%%%%%%%%%%%%%

\end{document}